\documentstyle[psfig]{mn2e}
\input epsf

\def\deg{\hbox{$^\circ$}}

\def\lesssim{\mathrel{\hbox{\rlap{\hbox{\lower4pt\hbox{$\sim$}}}\hbox{$<$}}}}
\def\gtrsim{\mathrel{\hbox{\rlap{\hbox{\lower4pt\hbox{$\sim$}}}\hbox{$>$}}}}

\begin{document}

\title[A WO star in the Scutum-Crux arm]
{Discovery of a WO star in the Scutum-Crux arm of the inner Galaxy}
\author[Janet E. Drew et al. ]
{Janet E. Drew$^1$, M. J. Barlow$^2$, Y. C. Unruh$^1$, Q. A. Parker$^{3,4}$, 
R. Wesson$^2$,
\newauthor M. J. Pierce$^5$, M. R. W. Masheder$^5$ \& S. Phillipps$^5$\\ 
$^1$Imperial College of Science, Technology and Medicine,
Blackett Laboratory, Exhibition Road, London,  SW7 2AZ, U.K.\\
$^2$University College London, Department of Physics \& Astronomy, 
Gower Street, London WC1E 6BT, U.K.\\
$^3$Department of Physics, Macquarie University, NSW 2109, Australia\\
$^4$Anglo-Australian Observatory, PO Box 296, Epping NSW 1710, Australia\\
$^5$Astrophysics Group, Department of Physics, Bristol University, 
Tyndall Avenue, Bristol, BS8 1TL, U.K.\\
}

\date{received,  accepted}

\maketitle
\begin{abstract}
We report the discovery of only the fourth massive WO star to be found in the 
Milky Way, and only the seventh identified within the Local Group.  This has 
resulted from the 
first observations made in a programme of follow-up spectroscopy of candidate 
emission line stars from the AAO/UK Schmidt Southern Galactic Plane H$\alpha$ 
Survey.  The optical spectrum of this star, to become WR~93b in the Catalogue
of Galactic Wolf-Rayet stars, is presented and described.  WR~93b is
classified as WO3 and is shown to be highly reddened ($E_{B-V} = 2.1\pm0.1$). 
A recombination line analysis of the emission lines yields the 
abundance ratios C/He = 0.95 and O/He = 0.13 (by number).
Comparisons at near infrared wavelengths of reddening corrected photometry
between WR~93b and both of Sand~2 (WO3, $D = 49$~kpc) and Sand~5 (WO2, $D = 
1.75$~kpc) yields a consistent distance to WR~93b of 3.4~kpc.  Positioned at
Galactic co-ordinates $\ell = 353.27^{\rm o}$, $b = -0.85^{\rm o}$, the star 
is most 
likely located in the Scutum-Crux Arm of the inner Milky Way. We note 
that none of the four Galactic WO stars lies significantly beyond the Solar 
Circle (with two well inside).

Estimation of the wind terminal velocity in WR~93b at 5750~km~s$^{-1}$ makes
this star the current wind speed record holder among all non-degenerate
stars.
\end{abstract}


\section{Introduction}

    In this paper we present the discovery of a particularly rare type of
emission line star made in the initial phase of a long-term spectroscopy 
programme which aims to confirm and provide preliminary classification of 
spatially-unresolved line-excess objects contained within
the AAO/UK Schmidt Southern Galactic Plane H$\alpha$ Survey (Parker et al 
2003).  The star in question has been revealed as only the fourth massive WO 
star to be found in the Galaxy -- bringing the total identified within the 
Local Group to just seven.  When it is assimilated into the next revision of 
the catalogue of massive Wolf-Rayet stars in the Galaxy, compiled and 
maintained by van der Hucht (2001), its position in the sky will earn it
the designation WR 93b.  Hereafter 
we refer to this newly discovered star by this name.  The WO stars are the 
most chemically extreme Wolf-Rayet (WR) stars, whose spectra are dominated by 
high-excitation oxygen and carbon lines.  Objects in this elite group of 
stars are viewed as plausible progenitors for extreme, chemically-peculiar 
Type Ib/c supernovae (Woosley, Heger \& Weaver 2002) and some GRBs (e.g. 
Schaefer et al 2003). 

The history of this presently rare type of object begins with
the last generation of galactic emission line surveys: Sanduleak (1971) 
presented 
a list of five WR stars, two in the Magellanic Clouds, which had 
strong O~{\sc vi} 3811,34~\AA\ emission features that had previously  
been found only amongst planetary nebula nuclei.  Barlow \& Hummer (1982) 
proposed that one of these five stars was indeed a PN central star but that 
the other four corresponded to an advanced stage of evolution of massive 
stars, beyond the WC phase, and classified them as WO Wolf-Rayet stars.
Since this time, two more have been added to this grouping (MS 4, 
by Smith, Shara \& Moffat 1990 and DR~1 in the dwarf irregular galaxy IC 1613 
by Kingsburgh \& Barlow 1995).
In general terms, discoveries of Galactic WR stars have been assisted greatly 
by optical objective-prism surveys (Stephenson \& Sanduleak 1971, MacConnell \&
Sanduleak 1970) and narrow-band imaging 
surveys (most recently, Shara et al. 1999). The result we present here uses 
the latter technique in the red part of the spectrum, and serves as an 
encouragement that much remains to be discovered.

In the next section we introduce the AAO/UK Schmidt Southern Galactic Plane
H$\alpha$ Survey and describe the data it contains relevant to the
newly-discovered WO star and its locale.  We then describe how the WO star
came to be included in our initial programme of follow-up spectroscopy
using the AAO/UK Schmidt multi-fibre spectroscopy facility, 6dF.  This is 
followed, in Section 3, by presentation
of both the 6dF data obtained in May/June 2003 (\S 3.1), and 
flux-calibrated long-slit spectra obtained at the William Herschel Telescope
(WHT) in August 2003 (\S 3.2).  We are then in a position
to classify the WO star and determine its reddening (\S 4).  We also
present estimates of C/He and O/He abundance ratios deduced from the
emission lines (\S 5).  Finally, by comparison with other WO stars
at known distances, we derive the distance to WR~93b (\S 6): at 
3.4~kpc, and along a line of sight passing within a few degrees of the 
Galactic Centre, it seems likely the new WO star is associated with the 
Scutum-Crux spiral arm (Russeil 2003) -- without a doubt it is well inside 
the Solar Circle. We close in \S 7 with comment on the 4-strong group of 
galactic WO stars and on the prospects for future discoveries.

\section{UKST H$\alpha$ survey imaging observations}

\begin{figure}
\begin{picture}(0,200)
\put(0,0)
{\includegraphics{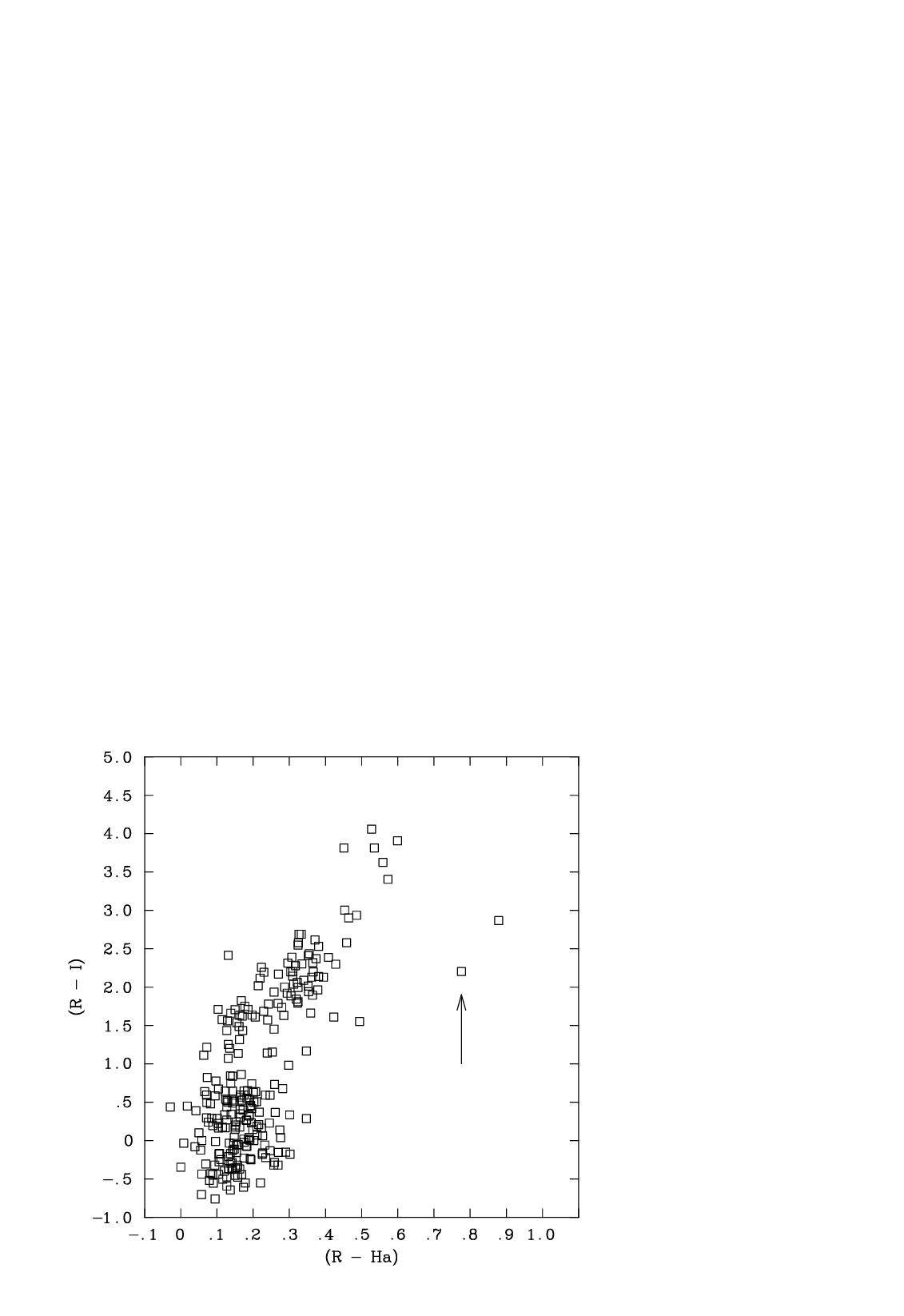}}
\end{picture}
\caption{A plot of $(R-I)$ versus $(R-H\alpha)$ for all SHS catalogue 
point sources with $14.5 < R < 16.0$, lying within a 15~arcmin radius of 
WR~93b.  The WO star itself is indicated by the arrow.
Note that most stars in this diagram congregate in the rectangle defined
by $(R-I) < 2$ and $-0.1 < (R-H\alpha ) < 0.3$: the stars trailing off in
a lightly sloping line to higher $(R-I)$ and $(R-H\alpha )$ are mainly 
extreme red stars with prominent molecular band structure in the red part
of the spectrum. Interstellar reddening will displace a star vertically
in this diagram. For the most part, the SHS database is calibrated to
yield $(R-H\alpha ) \simeq 0$ in the mean -- here, in this small field,
there happens to be an offset to between 0.1 and 0.2 in this colour. }
\label{f_cc}
\end{figure}

The AAO/UKST narrow-band H$\alpha$ Survey of the Southern Galactic Plane and
Magellanic Clouds is the last photographic sky survey to have been carried
out on the UK Schmidt Telescope (UKST). It has recently been completed
(Parker et al. 2003) and is now available as digital survey data
derived from SuperCOSMOS scans of the original survey films at 10~$\mu$m
(0.67 arcsec) resolution (hereafter, the SHS database, located at 
http://www-wfau.roe.ac.uk/sss/halpha/).  This survey used a specially 
designed, 70~\AA\ bandpass, H$\alpha$ interference filter (Parker \& 
Bland-Hawthorn 1998).  When combined with fine-grained Tech-Pan emulsion as 
detector (Parker and Malin, 1999), it made possible a survey of Galactic 
ionised hydrogen that is a powerful combination of depth, spatial resolution 
and areal coverage (Parker \& Phillipps, 1998).  Each field observed has an 
effective dimension projected on the sky of $4\deg \times 4\deg$.  The survey 
policy was to gather contemporaneous 15-minute broad-band `short $R$' 
exposures alongside 3-hour narrow-band H$\alpha$ exposures in order to yield 
exposure pairs, well matched both in terms of sensitivity limit ($R = 19.5$) 
for continuum 
point sources and in terms of point-spread functions. The entire southern 
plane of the Milky Way has been imaged within the Galactic latitude range 
$-10\deg < b < +10\deg $ in this way.   These exposure pairs provide the 
source material for a variety of continuing research projects (see e.g. 
Morgan, Parker \& Russeil 2001; Parker \& Morgan 2003; Parker et al 2003).

We have embarked on a programme aimed at providing spectroscopic
confirmations of spatially-unresolved candidate emission line stars
derived from the SHS database.  This began in 2003 with limited data-taking
using the UKST 6dF facility to follow up a few fields in the brighter half
of the available magnitude range ($12.5 < R < 16$), with a view to
establishing observing procedures for the longer term.  The broad goal of
the programme is to increase by an order of magnitude Galactic samples of 
the relatively short-lived, rare phases of stellar evolution typified by 
strong line emission at H$\alpha$.  This is expected to be achieved
by dropping the current typical limit of $R \sim 13$ on survey samples
of emission line stars (cf. MacConnell 1982, Robertson \& Jordan 1989) down 
to $R\sim 19$.  The value
in this is that several groups of young and evolved objects, critical
to our understanding of pre- and post-main-sequence stellar evolution, are 
currently extremely poorly sampled: as noted already, the massive WO star 
reported here is only the fourth to be found in the Milky Way.

As part of our initial spectroscopy programme, we chose to observe targets 
drawn from field HAL0555 within the magnitude range $14.5 < R < 16.0$. 
Centred at 17h 20m $-36^o$ (1950 co-ordinates), this field is a highly 
reddened one straddling the Galactic equator, just a few degrees in longitude 
away from the Galactic Centre. 106 point sources were selected for 
spectroscopy from a longer list of candidate stars with relatively extreme 
H$\alpha$ magnitude excesses with respect to the associated short-red 
broadband magnitudes ($(R - H\alpha) > 0.6$, with the great majority of stars 
falling in the range $-0.2 < (R - H\alpha) < 0.5$).  Note that the SHS 
H$\alpha$ magnitudes are derived from calibration of the H$\alpha$ images 
against $R$ standards, and undergo final adjustments as a function of $R$ 
magnitude such that the median $(R - H\alpha)$ colour for any given field is 
set to zero (cf. the discussion of calibration of the SuperCOSMOS Sky Survey
presented in Hambly et al 2001).  The plan for the spectroscopy, in the long 
term, is to also select on $(R - I)$ in order to reduce the number of unwanted 
red stars whose molecular band structure mimics H$\alpha$ excess.  However 
this was not done here so that we might determine empirically how best to 
apply such a cut.

One of the sources selected for spectroscopy was located in the dark NE
part of the field at RA~17 32 03.30, Dec~$-35$ 04 32.5 (J2000) and was listed 
in the SHS database as having $R = 14.672$, and $(R - H\alpha) = 0.776$ (see
Hambly et al 2001 for a discussion of likely photographic errors).
The suitability of this object for follow-up spectroscopy becomes apparent
on considering its position in a magnitude-limited plot of point-source 
$(R-I)$ colour versus $(R - H\alpha)$ excess for objects in its immediate 
locality.  In Figure~\ref{f_cc} we present such a plot for all stars within 
15~arcmin of the WO star:  it can be seen there that it is one of just two 
objects sitting in relative isolation, on the `excess' side of a well-populated
locus of stars.  The other clear excess object star still awaits 
spectroscopic follow-up.

\section{Spectroscopic observations}

\begin{table*}
\caption{Abbreviated log of observations}
\label{t_obs}
\begin{tabular}{lllllrll}

\hline
telescope/ & observation & UT start & grating & central & exposures & seeing &
additional information \\
spectrograph & date      &          &         & wavelength & $n\times t$ &  &\\
             &           &          &         & (\AA )     & (sec) & (arcsec) &
  \\
\hline
UKST/6dF & 2003/05/24 & 15:55 & 1516R & 6490 & 5x1200 &  1--2 & \\
         &            &       &       &      & 5x1200 &  &
interlaced offsets (RA,Dec arcsec:\\
         &            &       &       &      &        &  &
$+$10,$+$5; $+$10,$-$5; $-$10,$+$5; $-$10,$-$5 \\
         &            &       &       &      &        &  & $+$5,$-$10) \\
UKST/6dF & 2003/06/06 & 16:08 &  580V & 4780 & 7x1200 &  3--4 &  \\
WHT/ISIS & 2003/08/02 & 20:35 & R600B & 4282 &  2x120 &  1    &
  HZ 44 flux standard \\
         &            &       &       &      &        &  &  
  airmass 1.22, 1.23  slit 8.0~arcsec \\
         &            &       &       &      &        &  & slit PA 90 \\   
         &            & 20:35 & R316R & 6547 &  2x120 &  &
  HZ 44 flux standard  \\
         &            & 21:03 & R600B & 4282 &  2x900 &  & 
  WO star: airmass 2.37, 2.32  \\
         &            &       &       &      &        &  & slit 0.8 arcsec, PA 0 \\
         &            & 21:03 & R316R & 6547 &  2x900 &  & WO star  \\
\hline
\end{tabular}
\end{table*}

\subsection{UKST/6dF Observations}

\begin{figure}
\begin{picture}(0,280)
\put(0,0)
{\includegraphics{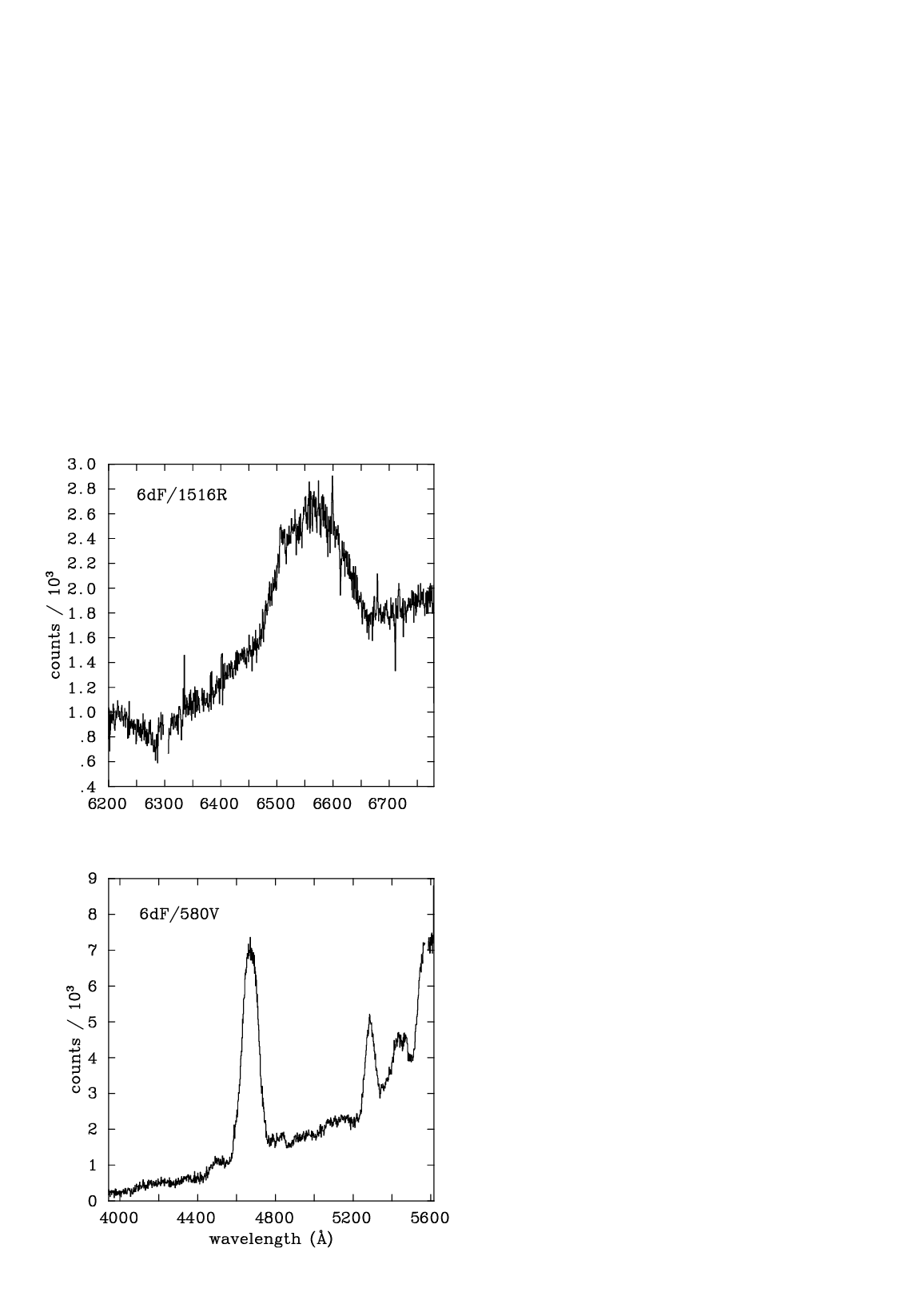}}
\end{picture}
\caption{The spectra of WR~93b obtained in May/June 2003
using 6dF.  The top panel contains the spectrum obtained using the 1516R
grating in the 'H$\alpha$' region (for this star, the emission at 
$\sim$6560~\AA\ is a He{\sc ii}/C{\sc iv} blend).  The bottom panel shows
the broader coverage 580V spectrum.}
\label{f_6df}
\end{figure}

Prompted as described above, we used the AAO/UK Schmidt 6dF multi-fibre 
spectrograph to obtain spectra of 106 spatially-unresolved objects located in 
field 555 of the imaging survey: every target selected satisfied the 
constraints $14.5 < R < 16$ and $(R - H\alpha) > 0.6$.  This target list was
observed first using the 1516R grating on 2003 May 24, and for a second
time on 2003 June 6 using the 580V grating (see Table~\ref{t_obs} for 
details).  The respective spectral resolutions of the two gratings are
$\sim 2$\AA\ and $\sim 6$\AA .  The 1516R observations were performed as 
a series of alternating on-target and sky-offset observations, to ensure
the best possible subtraction of diffuse H$\alpha$ emission by sampling 
nebulosity in the immediate environment of every target star.   



Both datasets were extracted from the CCD frames using the 6dF 
adaptation of the 2dfdr software package (see
http://www.aao.gov.au/AAO/2df/manual.htm).  The extraction
consisted of: bias subtraction using the overscan strip; flat field
extraction; fibre-by-fibre arc extraction and calibration; optimal
extraction of the observed spectrum from each fibre, followed by flat-field 
correction and wavelength calibration.  For the May 1516R observations,
the sky offset frames were each subtracted from the associated target 
frames and then combined, weighted according to their statistical quality, 
to give a single reduced frame 
containing up to 150 spectra each of 1032 wavelength-calibrated pixels.
No throughput correction was applied to these data.  The June 580V
observations were handled differently in that a fibre throughput
correction was applied so as to enable sky subtraction using the mean 
spectrum of the designated sky fibres.

The extracted 6dF spectra of WR~93b are shown in Figure~\ref{f_6df}.  It was 
immediately evident from these data that the object was a Wolf-Rayet star 
presenting extremely broad emission features.  Furthermore it was possible to 
note the partial capture of the O{\sc vi}~$\lambda$6200 line in the 1516R 
observation (Figure~\ref{f_6df}, top panel) and the clear presence of 
O{\sc vi}~$\lambda$5290 in the 580V observation (Figure~\ref{f_6df}, lower 
panel).  Either of these features at the observed prominence is sufficient 
to identify the object as a WO star.  It was also evident from both spectra, 
when considered in comparison to those of other targets, that 
WR~93b must be significantly reddened.  In order to obtain 
the extended blue coverage of the optical spectrum needed for WO sub-type 
determination (using ratios of emission line equivalent widths) and to 
assess the relative spectral energy distribution, it was decided to obtain 
additional observations using ISIS on the William Herschel Telescope.

\subsection{WHT/ISIS observations}

\begin{figure*}
\begin{picture}(0,305)
\put(0,0)
{\includegraphics{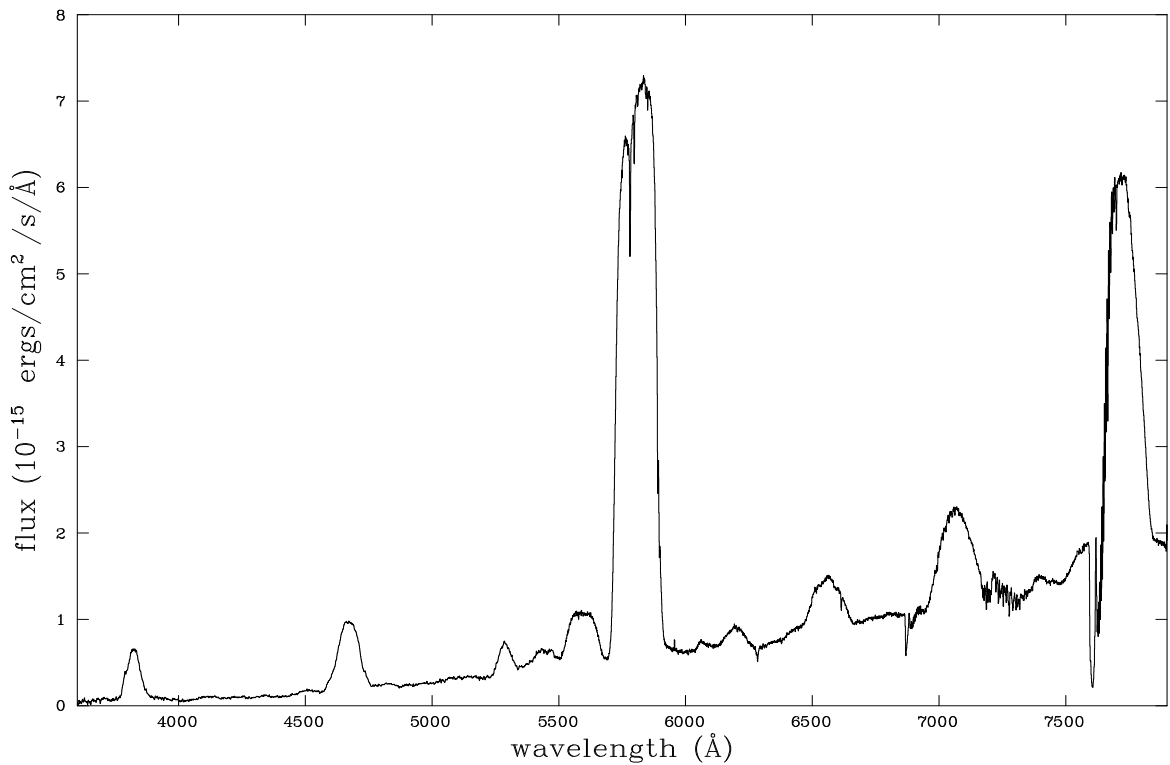}}
\end{picture}
\caption{The combined flux-calibrated spectrum of WR~93b 
obtained at the William Herschel Telescope on 2003 August 2, with the 6dF/580V
spectrum spliced in to cover 5000--5500~\AA .  Note that the absolute flux 
level plotted will be an under-estimate because of slit losses.}
\label{f_wht}
\end{figure*}

\begin{table*}
\caption{Emission line identifications and measurements from the
spectrum of WR~93b over the wavelength range 3600--7900~\AA . Observed 
wavelengths are centroid wavelengths to the nearest 5~\AA ; line widths 
(FWHM) are quoted to the nearest 10~\AA . A colon placed against a numerical 
value indicates greater uncertainty than for other values listed in the same 
column.  C~{\sc iv} $\lambda$7062 has some telluric absorption in its extreme
red wing. The last three emission lines listed are included for 
completeness: they were not measured in view of the more severe 
uncorrected telluric absorption overlapping them. Columns 4, 5 and 6
contain equivalent width (EW) measurements and the quantities derived from
them that are needed for the abundance determination presented in \S 5.
In the last two columns, and for the purpose of comparison with WR~93b,
equivalent widths are given for Sand~2 and and Sand~5 (data from KBS95). }
\label{t_lines}
\begin{tabular}{llllllll}
\hline
$\lambda_{obs}$ & ID & FWHM & EW & I & I/Q & 
      \multicolumn{2}{c}{Comparison EW (\AA )} \\ 
      & & & & (10$^{-12}$ ergs & &  Sand~2  & Sand 5 \\
(\AA ) & & (\AA ) & (\AA ) & cm$^{-2}$ s$^{-1}$) & (cm$^{-5}$) 
      & (WO3) & (WO2) \\ 
\hline
3820 & O~{\sc vi} 3811,34 (3p~$^2$P$\rightarrow$3s~$^2$S) & 60 & 670$\pm$100 &
      & & 336$\pm$6 & 1005$\pm$17 \\
4120: & O~{\sc v} 4119-4178 (3p$'$~$^3$D$\rightarrow$3s$'$~$^3$P$^o$), &  
      & 30$\pm$5 & & & & \\
      & O~{\sc v} 4120-4159 (3d$'$~$^3$P$^o$$\rightarrow$3p$'$~$^3$S) 
      & & & & & & \\
4505: & O~{\sc vi} 4499 (10$\rightarrow$8), & 60: & 23$\pm$5 & & & & \\
      & O~{\sc v} 4521 (9$\rightarrow$7), 
       4523 (3d$'$~$^1$P$^o$$\rightarrow$3p$'$~$^1$D) & & & & & & \\
4670  & C~{\sc iv} 4658 (6$\rightarrow$5), 4685 (8$\rightarrow$6), & 90 
      & 460$\pm$25 & & & 531$\pm$4 & 496$\pm$7 \\
      & He~{\sc ii} 4686 (4$\rightarrow$3) & & & & & & \\
4820: & ? & 50: & 10$\pm$5 & & & & \\
5285  & O~{\sc vi} 5290 (8$\rightarrow$7) & 50 & 58$\pm$3 & 11.0  
      &  $1.05\times10^{13}$ & 45$\pm$2 & 85$\pm$5 \\
5440: & He~{\sc ii} 5411 (7$\rightarrow$4), & 110: & 67$\pm$5 & & & 65$\pm$2 
      & 74$\pm$5 \\
      &  C~{\sc iv} 5411 (14$\rightarrow$8), 5470 (10$\rightarrow$7) & & & 
      & & & \\
5590  & O~{\sc v} 5590 (3d~$^3$D$\rightarrow$3p~$^3$P$^o$) & 130 & 170$\pm$7 
      & 27.4 & $6.3\times10^{12}$ & 112$\pm$3 & 222$\pm$10 \\
5805  & C~{\sc iv} 5801,12 (3p~$^2$P$^o$$\rightarrow$3s~$^2$S) & 160 & 
       1940$\pm$30 & 278 & $2.00\times10^{14}$ & 2450$\pm$190 & 374$\pm$10 \\
6070  & O~{\sc viii} 6068 (10$\rightarrow$9), 6064 (13$\rightarrow$11), & 60 
      & 9$\pm$3 & 0.87 & $4.3\times10^{11},$ & & \\
      & O~{\sc vii} 6085 (12$\rightarrow$10) & & & & $4.3\times10^{11}$ & & \\
6195  & O~{\sc vi} 6200 (11$\rightarrow$9, 13$\rightarrow$11) & 80 & 28$\pm$4 
      & 3.3 & $1.08\times10^{13}$ & 12$\pm$2 & 40$\pm$4 \\
6440  & O~{\sc v} 3d$'\rightarrow$3p$'$ & 70 & 9$\pm$3 & & & & \\
6555  & He~{\sc ii} 6560 (6$\rightarrow$4), & 120 & 85$\pm$7 & 8.5 & 
      $1.42\times10^{14}$ & 93$\pm$3 & 182$\pm$5 \\
      & C~{\sc iv} 6560 (12$\rightarrow$8) & & & & ($1.35\times10^{14}$) & & \\
7065  & C~{\sc iv} 7062 (9$\rightarrow$7) &  & 140$\pm$15 & 11.2 
      & $1.35\times10^{14}$ & 102$\pm$5 & 96$\pm$8 \\
7390  & O~{\sc v}? (8$\rightarrow$7 low $\ell$?) &  & weak & & & & \\
7590  & O~{\sc v} 7592, 7611 (8$\rightarrow$7) & & intermediate & & & & \\
7725  & C~{\sc iv} 7726 (7$\rightarrow$6), 7736 (11$\rightarrow$8),  & 
      & very strong & & & &  \\
      & O~{\sc vi} 7717 (9$\rightarrow$8) & & & & & & \\
\hline 
\end{tabular}
\end{table*}

Data were obtained on 2003 August 2nd using the twin-armed ISIS spectrograph
mounted on the William Herschel Telescope, located in La Palma.  It was a
clear night of good seeing. Because 
WR~93b is far south relative to the latitude of the
Isaac Newton Group on La Palma ($+28\deg$), care was taken to time the 
observations to within minutes of meridian transit, setting the slit angle to 
the north-south position so as to minimise slit losses caused by atmospheric 
dispersion.

The spectrograph setup included the 5400~\AA\ dichroic such that the blue
arm, with R600B grating, delivered a reliable spectrum over the range
3600--5000~\AA ; whilst the red arm with R316R grating gave good results
over the range 5500--7900~\AA .  We do not use data in the range 
5000--5500~\AA\ as these are visibly affected by the dichroic transmission
cut. The respective spectral resolutions obtained were 1.5~\AA\ and 2.8~\AA , 
giving e.g. a velocity 
resolution of 145 km~s$^{-1}$ at the C{\sc iv}~$\lambda\lambda$5801,12 
doublet.  Further details of the observations are given in Table~\ref{t_obs}.  
Wavelength calibration was performed using arc exposures obtained immediately 
after the stellar spectra, at the same telescope pointing.  Wide-slit
observations of the flux standard HZ~44 (Massey et al 1988) obtained a little 
earlier in the night were used to flux-calibrate the data after extraction to 
one-dimensional format.  Before this calibration was applied to the WO-star 
spectra, the necessarily large correction for air mass was also made.  The
ISIS spectra as displayed here have been smoothed using a gaussian of 
FWHM $=$ 1 \AA\ in order to achieve some noise reduction. No correction
has been made for telluric absorption.

In Figure~\ref{f_wht}, we show the flux-calibrated ISIS spectra together with 
a rescaled interpolation of the UKST/6dF 580V spectrum, that usefully spans
the 5000--5500~\AA\ gap left by the ISIS results.  It turns out that this
interpolation can be achieved with confidence because there is excellent 
agreement between the UKST and WHT observations in terms of both emission line
contrast and profile shape where the two overlap.  This is reassuring
as we would not expect, a priori, the WO-star line spectrum to vary between
the two epochs of observation.

\section{The spectral type, wind terminal velocity of, and reddening towards 
WR~93b}

   The outstanding properties of the merged flux-calibrated spectrum
(Figure~\ref{f_wht}) are the red slope of the continuum and the breadth and 
extraordinary prominence of the strongest emission lines.  For example, the 
continuum flux rises by a factor of $\sim10$ between 4000~\AA\ and 
$\sim$5700~\AA , whilst the C{\sc iv}~$\lambda\lambda$5801,12 emission stands, 
at its peak, at $\sim$13 times the continuum level (with an equivalent width 
of $\sim$1900~\AA !).   Observed wavelengths, likely identifications, FWHM
estimates and equivalent width measurements for the emission 
features seen in the spectrum are laid out in Table~\ref{t_lines}.

    In order to assign a spectral type to WR~93b, we refer to the 
classification criteria given by Crowther, de~Marco and Barlow (1998).  The 
primary discriminant is the ratio between the equivalent widths (EWs) of the 
O~{\sc vi}~$\lambda\lambda$3811,34 and O~{\sc v}~$\lambda\lambda$5590 features:
here this comes out at $3.9 \pm 0.6$, close to but just under the WO2/WO3 
boundary value of 4 (see data in Table~\ref{t_lines}, and Table 3 of Crowther 
et al 1998).  The secondary classification criteria are the 
(O~{\sc vi} $\lambda\lambda$3811,34/C~{\sc iv} $\lambda\lambda$5801,12) and 
(O~{\sc vii} $\lambda$5670/O{\sc v} $\lambda$5590) 
equivalent width ratios.  The value of the first of these is $0.35 \pm 0.05$, 
whilst the second ratio can only be described as small since the 
O~{\sc vii}~$\lambda$5670 cannot be said to have been definitely detected.  
These indicators are more clearly in line with a WO3 classification, which
we prefer for the time being.

This provisional classification places the new WO star in the same sub-type 
as the LMC WO star, Sand~2.  This similarity is reinforced on making a broader 
comparison between the equivalent width data (Table~\ref{t_lines}) given for 
WR~93b and reproduced for Sand~2 and the Galactic WO2 star, 
Sand~5, from Kingsburgh, Barlow \& Storey (1995, hereafter KBS95).  We can 
conclude that WR~93b exhibits slightly higher excitation than Sand~2, whilst 
certainly not being as extreme as Sand~5.  In due course, a more decisive 
spectral classification may become possible via a measurement of the relative 
equivalent widths of O~{\sc iv}~$\lambda$3400 (3d~$^2$D -- 3p~$^2$P$^o$) and 
O~{\sc vi} $\lambda\lambda$3434,38 (7--6,11--8): for the WO3 stars, Sand~1 
and Sand~2, KBS95 found EW(O~{\sc iv}) $>$ EW(O~{\sc vi}) whilst for the WO2 
stars Sand~4 and Sand~5, the O~{\sc iv} line was undetectable.

    To estimate the terminal stellar wind velocity, we measure the full width 
at zero intensity (FWZI) of the C~{\sc iv} $\lambda\lambda$5801,12 doublet. We 
adopt the same approach to this as did KBS95.
For the case of the SMC WO3 star, Sand~1, they showed that the velocity 
corresponding to half the FWZI of this bright optical line (after correction 
for the 10.65~\AA\ doublet separation) matched closely the terminal wind 
velocity measured from the black absorption edge of the C~{\sc iv} 1548,51~\AA\
resonance doublet in a high resolution {\em IUE} spectrum.  Respectively, 
these velocities were 4150 and 4200 km~s$^{-1}$.  For Sand~2, KBS95 derived a 
velocity of 4450~km~s$^{-1}$ from the C~{\sc iv} 5801,12~\AA\ FWZI/2, after 
correction for the doublet separation.  Crowther et al. (2000) subsequently 
obtained {\em HST} ultraviolet spectra of Sand~2 which yielded a wind terminal 
velocity of 4100~km~s$^{-1}$ from the black absorption edge of the 
C~{\sc iv} $\lambda\lambda$1548,51 doublet. For WR~93b, the 
measured FWZI of the C~{\sc iv} 5801,12~\AA\ doublet is 233.4~\AA . After 
taking off 10.65~\AA\ for the doublet separation, this feature's FWZI/2 yields 
a wind terminal velocity of 5750~km~s$^{-1}$ -- somewhat higher than the 
5500~km~s$^{-1}$ measured by KBS95 for the Galactic WO2 star Sand~5.  This
is the highest wind terminal velocity so far determined for any star.  
Certainly the measured FWHM of this very bright line (160~\AA ) 
is larger for WR~93b than for any of the other WO stars.

    Given that the star is a WR star, it is evident that we should expect 
its intrinsic SED to be very blue.  In this spirit, KBS95 estimated WO star 
reddenings using the finding that a wide range of Wolf-Rayet sub-types yield 
optical SEDs closely following the power law $F_{\lambda} \propto 
\lambda^{-3}$ (Morris et al 1993).  In the specific instances of Sand~2 in 
the LMC and also Sand~5, both with independent measures of reddening, KBS95 
were able to demonstrate the utility of this approach by finding that the 
dereddened SEDs did indeed fit with expectation (the best fit indices were 
respectively $-$3.04 and $-$3.02).  Accordingly, for WR~93b, 
we identify the range in $E(B-V)$ that restores the optical SED most nearly 
to a $\lambda^{-3}$ power law: in the case of a standard Galactic reddening 
law with $R = 3.1$, this is $2.0 < E(B-V) < 2.35$, or equivalently $6.1 <A_V 
< 7.2$.  

    Confidence in the reddening estimate grows on noting that corrections 
consistent with the deduced range have the added virtue of yielding 
convincingly flat-topped C~{\sc iv} $\lambda\lambda$5801,12 profile.  Before 
correction (Figure~\ref{f_wht}), there is a very marked red skew of the 
peak flux in the profile.  Such a strong line, formed by a relatively 
lowly-ionised species in the context of these stellar winds, can be presumed 
to sample the terminal flow very well.  The profile of such a line would be 
expected to have a reflection-symmetric, nearly `rectangular', appearance 
(interpolating across the tops of the superposed narrow interstellar 
absorption lines).  In fact, we find that the reflection symmetry is most 
perfect for $E(B-V) \simeq 2.1$, whilst at $E(B-V) \sim 2.3$, it begins to 
look as though the blue half of the line profile has been pushed high relative 
to the red half.

\begin{figure}
\begin{picture}(0,150)
\put(0,0)
{\includegraphics{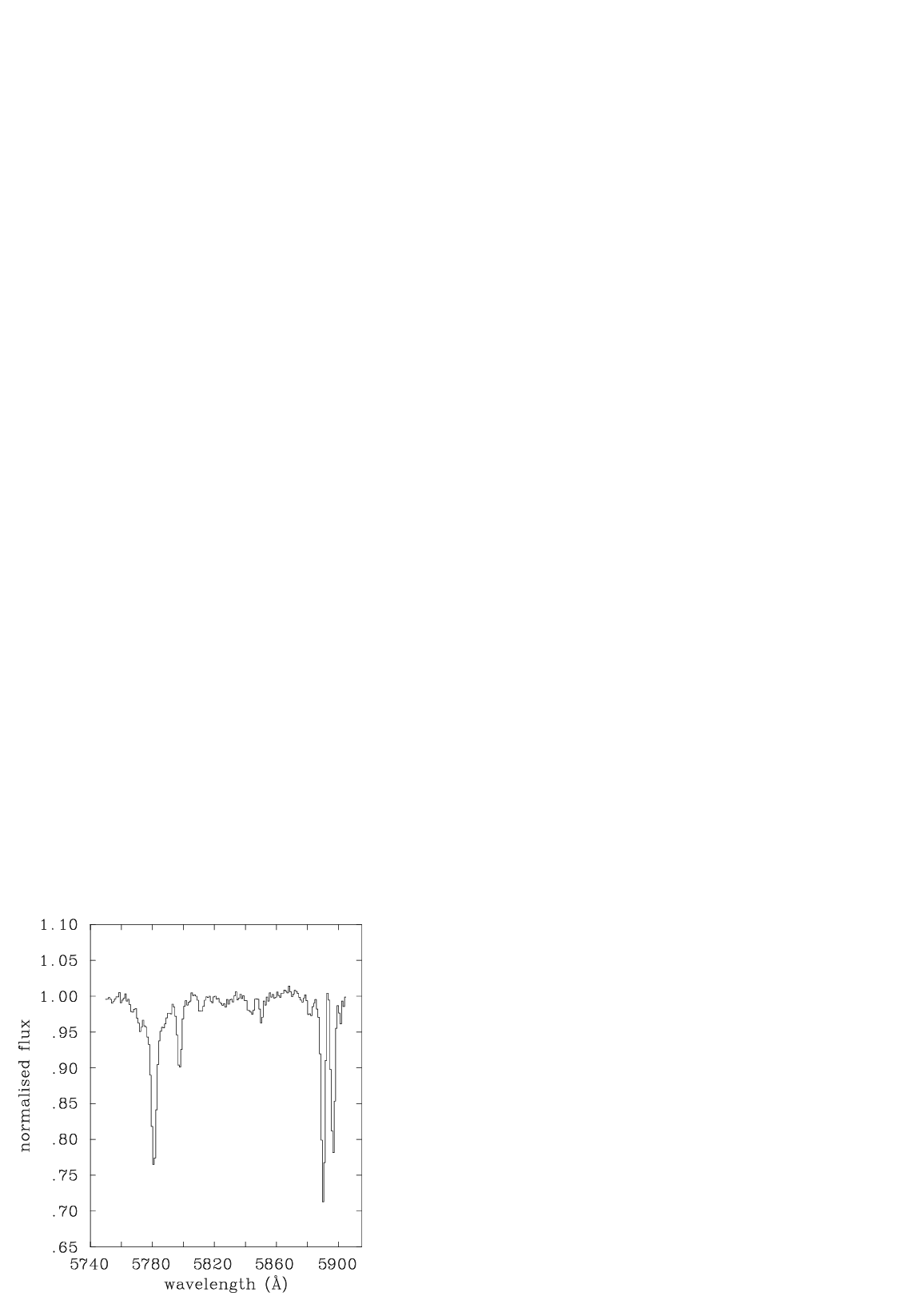}}
\end{picture}
\caption{The diffuse interstellar bands and Na~{\sc i} D lines seen at
high signal-to-noise ratio across the top of the 
C~{\sc iv}~$\lambda\lambda$5801,12 emission line profile. The spectrum
has been rectified by fitting out the shape of the C~{\sc iv} emission
profile (see figure~\ref{f_wht}). }
\label{f_dibs}
\end{figure}

\begin{table}
\caption{Equivalent width fits to DIBs located in the bright
C~{\sc iv} $\lambda\lambda$5801,12 emission, compared with results for the 
DIB reference star HD 183143 (Herbig 1995). Where no FWHM value is given it 
was set equal to the spectral resolution in the fit procedure.  The last 3 
rows also list equivalent width measurements for the well-known interstellar 
Na~{\sc i} and K~{\sc i} absorption lines: the reference wavelengths are 
laboratory values in these cases.}
\label{t_dibs}
\begin{tabular}{llrrr}
\hline
$\lambda_{ref}$ & $\lambda_{obs}$ & FWHM & EW & EW$_{ref}$ \\ 
(\AA ) & (\AA ) & (\AA ) & (m\AA ) & (m\AA ) \\
\hline
5766.05 & 5766.9$\pm$0.7  &              &    38$\pm$23 &  39 \\  
5772.49 & 5771.8$\pm$0.5  &              &    60$\pm$25 &  31 \\ 
5778.3  & 5781.2$\pm$0.6  & 17.5$\pm$2.3 &  1061$\pm$80 & 950 \\
5780.45 & 5781.0$\pm$0.05 & 3.4$\pm$0.2  &   673$\pm$52 & 801 \\
5796.98 & 5797.6$\pm$0.1  & 3.3$\pm$0.3  &   347$\pm$30 & 238 \\
5809.13 & 5811.5$\pm$0.45 & 3.5$\pm$1.0  &    86$\pm$23 &  34 \\
5843.6  & 5843.5$\pm$0.5  & 4.5$\pm$1.1  &   122$\pm$26 & 118 \\
5849.65 & 5850.4$\pm$0.25 &              &   100$\pm$17 &  82 \\
\hline
5889.95 & 5890.2$\pm$0.05 &              &   859$\pm$17 &   \\
5895.92 & 5896.5$\pm$0.05 &              &   670$\pm$17 &   \\
7698.98 & 7699.3$\pm$0.1  &              &   250$\pm$20 &   \\
\hline
\end{tabular}
\end{table}

\begin{table*}
\caption{Comparison between 2MASS broadband $J$, $H$ and $K$ magnitudes for 
WR~93b and those of the LMC WO3 star, Sand~2 and Galactic WO2
stars, Sand~5. The letter in parentheses after each observed magnitude is
the quality flag specified in the 2MASS database.}
\label{t_IR}
\begin{tabular}{llllllllll}
\hline
 & & \multicolumn{2}{l}{WR~93b} & \multicolumn{2}{l}{Sand~2} & \multicolumn{2}{l}{Sand~5} 
& \multicolumn{2}{l}{corrected magnitude} \\
 & & \multicolumn{2}{l}{($E(B-V) = 2.1\pm 0.1$)} & \multicolumn{2}{l}{($E(B-V) = 0.19\pm 0.03$)} &
   \multicolumn{2}{l}{($E(B-V) = 1.7\pm 0.1$)} & \multicolumn{2}{l}{differences} \\
  & $A_{\lambda}/A_V$ & observed & corrected & observed & corrected & observed & corrected & (6)-(4) & (8)-(4) \\
\hline
$J$ & 0.282 & 11.33$\pm$0.03 (A) & 9.49$\pm$0.10  &  15.44$\pm$0.06 (A) & 15.27$\pm$0.07  & 9.54$\pm$0.02 (A) & 8.05$\pm$0.09 & 5.78$\pm$0.12 & -1.44$\pm$0.13  \\ 
$H$ & 0.190 & 10.56$\pm$0.04 (A) & 9.32$\pm$0.07  &  15.33$\pm$0.13 (B) & 15.22$\pm$0.13  & 8.89$\pm$0.02 (A) & 7.89$\pm$0.06 & 5.90$\pm$0.15 & -1.43$\pm$0.09  \\
$K$ & 0.114 & 10.17$\pm$0.04 (A) & 9.43$\pm$0.06  &  14.88$\pm$0.14 (B) & 14.82$\pm$0.14  & 8.60$\pm$0.02 (A) & 8.00$\pm$0.04 & 5.39$\pm$0.15 & -1.43$\pm$0.07  \\
\hline
  & & & & & \multicolumn{3}{r}{adopted mean differences:} & 5.7$\pm$0.10 & -1.43$\pm$0.04 \\
\hline
\end{tabular}
\end{table*}

    A further property of the observed spectrum is the presence of
diffuse interstellar band (DIB) absorption.  For example, arrayed very clearly
across the top of the very same C{\sc iv}~$\lambda\lambda$5801,12 feature 
are a number of DIBs, together with the interstellar Na~{\sc i} D lines
(see Figure~\ref{f_dibs}) perched on its redward side.  We gain some insight 
into the physical character of the sightline to WR~93b if we 
compare the equivalent widths of these DIBs with their counterparts observed 
in the spectrum of HD~183143, a reference object for DIB studies (Herbig 1995).
These data are laid out in Table~\ref{t_dibs}.  It can be seen that the two 
sets of equivalent widths are broadly comparable, despite the fact the 
accepted colour excess for HD~183143 is $E(B-V) = 1.3$, significantly
less than determined from the SED of the WO star. (Note, in particular, 
that the fitting procedure, which had difficulty locating and distinguishing
the broad 5778~\AA\ DIB beneath the much sharper 5780.45~\AA\ DIB, returns an 
equivalent width sum for these two features that differs negligibly from that 
measured for HD~183143.)  This discrepancy amounts to a strong indication that 
a substantial part of the WO star reddening arises in intervening molecular 
gas, rather than solely in diffuse-cloud atomic gas with which the DIB column 
is known to correlate (see Herbig 1995 and references therein).  

\section{Estimation of the C/He and O/He abundance ratios from the emission
line spectrum}

We have used the recombination line method of KBS95 (their Section 8.2) to
derive the relative abundances of helium, carbon and oxygen ions in the
wind of WR~93b. We used the emission line equivalent widths listed in
column 4 of Table~\ref{t_lines} to calculate the dereddened line fluxes, $I$, 
in column 5, by assuming that the underlying continuum follows a F$_{\lambda}
\propto \lambda^{-3}$ spectrum (normalised at 6000~\AA\ to
1.30$\times10^{-13}$ ergs~cm$^{-2}$~s$^{-1}$~\AA $^{-1}$).  The adopted
line emission coefficients, $Q$ $(= h\nu \alpha_{rec})$, were taken from
Table~14 of KBS95.  The ratio $I/Q$, given in column 6 of Table~\ref{t_lines} 
is linearly proportional to ion abundance.

It is safe to assume that the overwhelmingly dominant ion stage of carbon
in the wind of WR~93b is C$^{4+}$, and that helium is ionized to He$^{2+}$.  
Following KBS95, we use the isolated C~{\sc iv} 9-7 7062~\AA\ line as the 
preferred diagnostic for the C$^{4+}$ abundance, to obtain an $I/Q$ of 
1.35$\times10^{14}$ cm$^{-5}$. We used this $I/Q$ to then estimate the 
fraction (45\%) of the 6560~\AA\ feature flux contributed by C~{\sc iv} 12-8 
and 7p-5s. The remaining 55\% of the 6560~\AA\ flux could then be attributed 
to He~{\sc ii} 6-4: on this basis, an $I/Q$ of 1.42$\times10^{14}$ cm$^{-5}$ 
was derived for He$^{2+}$. 

Unlike carbon, oxygen can be expected to be and indeed is observed to be
distributed across a number of ion stages.  The isolated O~{\sc vi} lines at 
5290~\AA\ (8-7) and 6200~\AA\ (11-9,13-11) yielded reassuringly similar $I/Q$ 
values for O$^{6+}$ of 1.05$\times10^{13}$ cm$^{-5}$ and 1.08$\times10^{13}$ 
cm$^{-5}$, respectively.  For the other recombining oxygen ion seen 
prominently 
in the spectrum, O$^{5+}$, we use the O~{\sc v} 5590~\AA\ 3d-3p emission to 
determine an $I/Q$ of 6.3$\times10^{13}$ cm$^{-5}$.  Finally, in order to fit
the weak, very high excitation, 6070~\AA\ feature we assumed, like KBS95, 
that the abundances of (and hence $I/Q$ values for) O$^{7+}$ and O$^{8+}$ were 
the same, on which basis the O~{\sc vii} 12-10 transition at 6085~\AA\ is 
found to contribute 15\% of the total flux, with the O~{\sc viii} 10-9, 13-11 
lines at 6068,6064~\AA\ contributing the remainder:  for both ions,  
$I/Q = 4.3\times10^{11}$ cm$^{-5}$.

Summing all ionic species, we find C/He = 0.95, O/He = 0.13 and (C+O)/He =
1.08, by number. The O/He ratio of 0.13 derived for WR~93b is similar to
the values (0.10-0.11, by number) derived by KBS95 for the LMC WO3 star
Sand~2 and for the Galactic WO2 stars Sand~4 and Sand~5. However, the C/He
ratio of 0.95 for WR~93b is sigificantly higher than the C/He ratios of
0.51-0.52 derived for these other WO stars.  Expressed as mass fractions
the abundances obtained here for WR~93b are X(He) = 0.23, X(C) = 0.65 and 
X(O) = 0.12.

\section{The distance to WR~93b}

    A means to estimating the distance to the newly-discovered
WO star is to apply the estimated total visual extinction in combination with 
a scaling to the reddening-corrected magnitudes of the WO3 star, Sand~2, 
and WO2 star, Sand 5.  For both comparison stars -- so close in spectral type
to that of the new WO star -- reliable reddenings and 
distances can be derived from the literature: Sand~2 is in the LMC with a 
modest reddening of $E(B-V) = 0.19$ (Crowther et al 2000), whilst Sand~5 is 
located in the open cluster Berkeley 87, in the Cygnus X region of the Galaxy, 
behind a reddening of $E(B-V) = 1.7\pm 0.1$ (Turner \& Forbes 1982).  We must 
assume, of course, that WR~93b, Sand~2 and Sand 5 share similar 
intrinsic stellar and spectral properties.  It is an encouraging start that
the continuum $M_V$s for these references objects are indeed almost the same: 
for Sand~2, $M_V = -3.0$ (Crowther et al 2000), whilst for Sand~5, 
$M_V = -3.1$ for $V = 13.37$ (KBS95) at 1.75~kpc (see below).  

    To minimise 
the impact of reddening uncertainties on this comparison, it can be performed 
in the near-infrared using data taken from the 2MASS database (see 
http://www.ipac.caltech.edu/2mass/).  The 2MASS $JHK$ 
magnitudes for the three stars are set out in Table~\ref{t_IR}.  Using the 
tabulation of $A_{\lambda}/A_V$ given by Cardelli, Clayton \& Mathis (1989) 
and the available visual extinctions we correct the NIR magnitudes to zero 
reddening.  A little experimentation with the value of $E(B-V)$ for
WR~93b reveals that $E(B-V) \simeq 2.1$ is again favoured
as it yields dereddened NIR magnitudes that are most uniformly offset
from the high quality photometry of Sand~5. The consistency of outcome with 
respect to Sand~2 for this same reddening is also satisfactory given the 
larger photometry errors reported for this object.  We find, after applying
the reddening corrections, that the new WO star is $\sim$5.7 magnitudes 
brighter than Sand~2, and fainter than Sand~5 by $\sim$1.43 magnitudes. 

\begin{table*}
\caption{The names, spectral types, Galactic co-ordinates and distances to the
known Galactic WO stars, incorporating the newly discovered WO star,
WR~93b.  The spectral types for all but WR 93b are from Crowther et al (1998):
that for WR93b has been estimated here using the same criteria.
Two distances are given: the first ($D_{\odot}$) is the 
derived distance from the Sun, while the second ($D_{GC}$) is the distance 
from the Galactic Centre.  The calculation of the latter assumes the Sun is at 
a radius of 7.94~kpc (Eisenhauer et al 2003) and neglects the small angular 
displacement of each WO star out of the Galactic equatorial plane.  The 
references in the final column are the sources for the heliocentric distance
in each case.} 
\label{t_WOstars}
\begin{tabular}{lllrrrrl}
\hline
Object & WR catalogue & spectral type & \multicolumn{2}{r}{Gal. co-ords.} 
& $D_{\odot}$ & $D_{GC}$ & Distance reference \\ 
  & number &  & $\ell \deg $ & $b\deg $ & (kpc) & (kpc) &  \\
\hline
MS 4 & WR 30a & WO4+O5 & 288.90 & -1.38 & 7.8 & 9.2 & van der Hucht 2001\\
     & WR 93b & WO3 & 353.27 & -0.85 & 3.4 & 4.6 & this paper\\
Sand 4 & WR 102 & WO2 & 2.38 & +1.41 & 4.6 & 3.4 & this paper\\
Sand 5 & WR 142 & WO2 & 75.73 & +0.30 & 1.8 & 7.7 & Be 87 cluster: Massey et 
al 2001, \\
       &        &     &       &       &     &     & Kn\"odlseder et al 2002 \\
\hline
\end{tabular}
\end{table*}

\begin{table*}
\caption{Summary list of the parameters of WR 93b}
\label{t_summary}
\begin{tabular}{lll}
\hline
Parameter & value & comment\\ 
\hline
Position (RA,Dec) & 17 32 03.30 $-$35 04 32.5 & J2000, from SHS database \\
$R$, $I$ magnitudes & 14.7, 12.5 & from SHS database\\
$J$, $H$, $K$ magnitudes & 11.33$\pm$0.03, 10.56$\pm$0.04, 10.17$\pm$0.04 &
   from 2MASS database \\
E($B-V$) & 2.1$\pm$0.1 & $R=3.1$ gives A$_V \simeq 6.5$  \\
spectral type & WO3 & criteria place WR 93b closer \\
              &     & to WO2/3 boundary than Sand~2 \\
wind terminal velocity & 5750 km s$^{-1}$ & \\
abundance ratios: & C/He $=$ 0.95 & by number \\
                  & O/He $=$ 0.13 & \ `` \ \ ``     \\
                  & (C+O)/He $=$ 1.08 & \ `` \ \ ``     \\
distance      &  3.4$\pm$0.3 kpc & \\
\hline
\end{tabular}
\end{table*}

    Adopting a distance of 49~kpc to the LMC (see e.g.Gibson 2000), the 
difference 
of 5.7 magnitudes with respect to Sand~2 yields a distance of 
3.5~kpc to WR~93b. Two recent estimates of the distance to 
Be~87, the cluster hosting Sand~5, are respectively 1.6~kpc (Massey, 
DeGioia-Eastwood \& Waterhouse 2001) and 1.9~kpc (Kn\"odlseder et al 2002): 
we adopt the intermediate value of 1.75~kpc.  The NIR magnitude difference 
of $-$1.43, combined with this distance then yields a distance to 
WR~93b of 3.4~kpc.  This is very
satisfying agreement. But, given the systematic uncertainty contained 
within the assumption of shared stellar parameters, we should anticipate
an error on 3.4~kpc of up to 10 \% (equivalent to $\sim$0.2 offset in
$M_V$).
 
    The Galactic co-ordinates of WR~93b are $\ell = 353.27\deg$,
$b = -0.85\deg$, placing it less than 7\deg from the Galactic Centre line of
sight, and outside the strips surveyed for Wolf-Rayet stars by Shara et al 
(1999).  At a distance of 3.4~kpc the new WO star would be located very close 
to the structure Russeil (2003) identifies as the Scutum-Crux arm, well inside 
the long-established Sagittarius-Carina arm at $\sim$2~kpc in the same 
direction.  For a  necessarily young object like a Wolf-Rayet star this is a 
perfectly acceptable place to be. Projected onto the plane of the sky, the 
nearest H~{\sc ii} region (a signature of continuing star formation) listed by 
Russeil (2003, Table 3) is about half a degree away from WR~93b at 
$\ell = 353.43\deg$, 
$b = -0.368\deg$.  At 3.4~kpc this angular separation converts to a length of 
about 30~pc.  The kinematic distance to this H{\sc ii} region, which is 
`compact' and without a known optical counterpart, has been estimated from 
Caswell \& Haynes' (1987) measurement of its radio recombination line radial 
velocity.  In the framework adopted by Russeil (2003), wherein the Galactic 
Centre is 8.5~kpc away, the distance to this H~{\sc ii} region is required to 
be $3.5^{+0.6}_{-0.7}$ kpc.  This is consistent with our estimate of the 
distance 
to the WO star. If the Galactic Centre is taken to be 7.94$\pm$0.42~kpc as 
determined recently by Eisenhauer et al (2003), this revises the kinematic
distance downwards a little to 3.3~kpc -- leaving the quality of agreement 
unchanged.

\section{Discussion} 

    With the addition of this newly-revealed WO star, the elite club of known 
Galactic WO stars has increased its membership to four.  We consider 
briefly how these stars are distributed within the disk of the Milky Way.  
In Table~\ref{t_WOstars} we list the Galactic co-ordinates and estimated 
distances of all 4 stars and use these data to calculate their Galacto-centric 
distances. The distance from the Sun to MS 4 (WR 30a) is adopted from 
van der Hucht (2001), which in turn is based on the absolute calibration of 
O stars due to Vacca, Garmany \& Shull (1996).  The distance to Sand~4 (WO2, 
also known as WR 102) is particularly uncertain, with literature estimates 
ranging from 2.3~kpc (Kingsburgh et al 1995) up to 5.6~kpc (van der Hucht 
2001).  Here we have derived and use a value of 4.6~kpc on the basis of the 
same IR photometric scaling relative to Sand 5 (also WO2) that we have used to 
determine the distance to WR~93b.  This object joins 
WR~93b in lying far inside the solar circle, whilst the other 
two WO stars, Sand~5 (WR 142) and MS 4 lie at Galactocentric radii similar to
that of the Sun. 

    The existence of this small grouping, taken on its own, would not 
encourage the idea that evolution to a longer-lived WO stage is favoured in 
regions of reduced metallicity (Maeder 1991).  Nevertheless it is a fact that 
the three WO stars located in other galaxies are all found in low metallicity 
environments (the LMC, SMC and IC~1613), as predicted by stellar evolutionary 
models such as those of Maeder (1991). If theory is qualitatively correct, 
the conclusion must be that a Milky Way WO star population at 
large Galactocentric radii may still be waiting to be discovered.  This 
remains a distinct possibility because the northern Galactic Plane, which 
includes the anti-centre region has not yet been surveyed systematically for 
Wolf-Rayet stars of any sub-type.  By contrast the southern Plane has been 
better served, both by the Stephenson \& Sanduleak (1971) objective prism 
survey and, most recently, by Shara et al (1999) -- and of course by the 
AAO/UKST H$\alpha$ survey exploited here.  

     An alternative phenomenological view of this still modest Galactic WO 
star population is that, as higher excitation, extreme examples of early WC 
stars (see e.g. Crowther et al 1998), they should follow a similar trend with 
respect to Galactocentric radius to that exhibited by their less elusive 
relations.  If it should turn out that the WO stars are more frequent inside 
the solar circle than outside it, this would amount to the same trend as is 
now recognised for WC stars by e.g. Massey \& Johnson (1998) and  
Shara et al (1999).

     In this paper we have reported the bare facts on a new massive 
Galactic WO3 star (the only example of its sub-type in the Milky Way -- they 
are so few).  Our quantitative findings are summarised in 
Table~\ref{t_summary}.  It seems that this is a star with a record-holding 
stellar wind: no other wind from a non-degenerate star in the Galaxy is seen
to attain so high a terminal velocity (5750 km~s$^{-1}$).  It is also the 
most highly reddened WO star discovered so far, with $A_V \simeq 6.5$.   Last 
we note that WR~93b is remarkable also with regard to its carbon abundance 
relative to helium: no other known Wolf-Rayet star can match C/He = 0.95 by 
number.  Among the WO stars, only Sand~1 in the SMC approaches such an extreme 
ratio with C/He = 0.81 (KBS95).  But there the similarity ends in that, in 
Sand~1, O/He is also significantly enhanced relative to the other WO stars 
(O/He = 0.3, as compared with $\sim$0.1), while 
it is not in WR~93b.  As is often the case in the presence of such a sparse
sample of objects, the emergent picture from -- in this instance -- the known 
abundance patterns is too poorly formed to allow any astrophysical conclusions 
to be drawn.  The clear need is to find more of them.  We end on the 
optimistic note, that relative to the power of today's spectroscopic 
capabilities, WR 93~b at $R \simeq 14.7$ is a bright object: appropriate
spectroscopic surveys of the Galactic Plane are bound to find more.

\section*{Acknowledgments}
We would like to thank Nigel Hambly for valuable help on many occasions in 
connection with the exploitation of the SHS database.  MP and RW both 
acknowledge the support of postgraduate studentships funded by the Particle 
Physics \& Astronomy Research Council of the United Kingdom.  This paper
makes use of: data obtained with the AAO/UK Schmidt Telescope at Siding Spring
Observatory, NSW, Australia; data from the William Herschel Telescope, 
operated on the island of La Palma by the Isaac Newton Group in the Spanish 
Observatorio del Roque de los Muchachos of the Instituto de Astrofisica de 
Canarias; data products from the Two Micron All Sky Survey, which is a joint 
project of the University of Massachusetts and the Infrared Processing and
Analysis Center/California Institute of Technology, funded by the National 
Aeronautics and Space Administration and the National Science Foundation.

\end{document}